\documentclass{ws-procs9x6}
\begin{document}

\title{Transverse Imaging of the Proton in
Exclusive Diffractive pp Scattering}
\author{C.~E.~Hyde-Wright$^{1}$, L.~Frankfurt$^{2}$, M.~Strikman$^{3}$,
C.~Weiss$^{4}$} 
\address{$^{1}$Old Dominion University, Norfolk, VA 23529, USA \\
$^{2}$School of Physics and Astronomy, Tel Aviv University,  
Tel Aviv, Israel\\
$^{3}$Dept.\ of Physics, Pennsylvania State U., University Park, 
PA 16802, USA\\
$^{4}$Theory Center, Jefferson Lab, Newport News, VA 23606, USA}

\maketitle

\abstracts{In a recent paper we describe a new approach 
to rapidity gap survival (RGS) in the production of high--mass systems
($H$ = dijet, Higgs, {\it etc.}) in exclusive double--gap diffractive $pp$ 
scattering, $pp \rightarrow p + H + p$. It is based on the idea that 
hard and soft interactions are approximately independent (QCD factorization), 
and allows us to calculate the RGS probability in 
a model--independent way in terms of the gluon generalized parton 
distributions (GPDs) in the colliding protons and the $pp$ elastic 
scattering amplitude. Here we focus on the transverse momentum dependence 
of the cross section. By measuring the ``diffraction pattern,'' one can 
perform detailed tests of the interplay of hard and soft interactions, and 
even extract information about the gluon GPD in the proton from the data.}

Production of high--mass systems ($H =$ dijet, diphotons, heavy 
quarkonium, Higgs boson, {\it etc.}) in exclusive double--gap
diffractive $pp$ scattering,
\begin{equation}
pp \;\; \rightarrow \;\; p \, + \, \text{(gap)} \, + \, H \, + \, 
\text{(gap)} \, + \, p,
\label{exclusive_diffraction}
\end{equation}
is of importance both as a promising channel for the Higgs boson search 
at the LHC\cite{Martin:2006fx}, and as a laboratory for studying 
strong interaction dynamics in high--energy collisions.
Diffractive events arise as the result of a
delicate interplay of hard and soft interactions. The high--mass 
system is produced in a hard scattering process, involving the exchange 
of two gluons between the protons; the requirement of the absence of
QCD radiation ensures the localization of this process in space and time.
In addition, one must require that the soft interactions between the 
spectator systems not lead to particle production. This results in 
a suppression of the cross section compared to the estimate based 
on the hard process alone, the so--called rapidity gap survival 
(RGS) probability\cite{Bjorken:1992er}.

In a recent paper\cite{RGS} we describe a new approach to RGS in
exclusive double--gap diffraction (\ref{exclusive_diffraction}) 
based on the idea that hard and soft interactions proceed over 
very different time-- and distance scales and are thus
approximately independent (QCD factorization). 
We show that the amplitude can be calculated in a model--independent 
way in terms of the gluon generalized parton 
distributions (GPDs) in the colliding protons and the $pp$ elastic
scattering amplitude; excitation of inelastic intermediate states 
is suppressed. In these proceedings we focus on the
transverse momentum dependence of the diffractive cross section.

%
%
\begin{figure}[b]
\centerline{\epsfxsize=0.7\textwidth \epsfbox{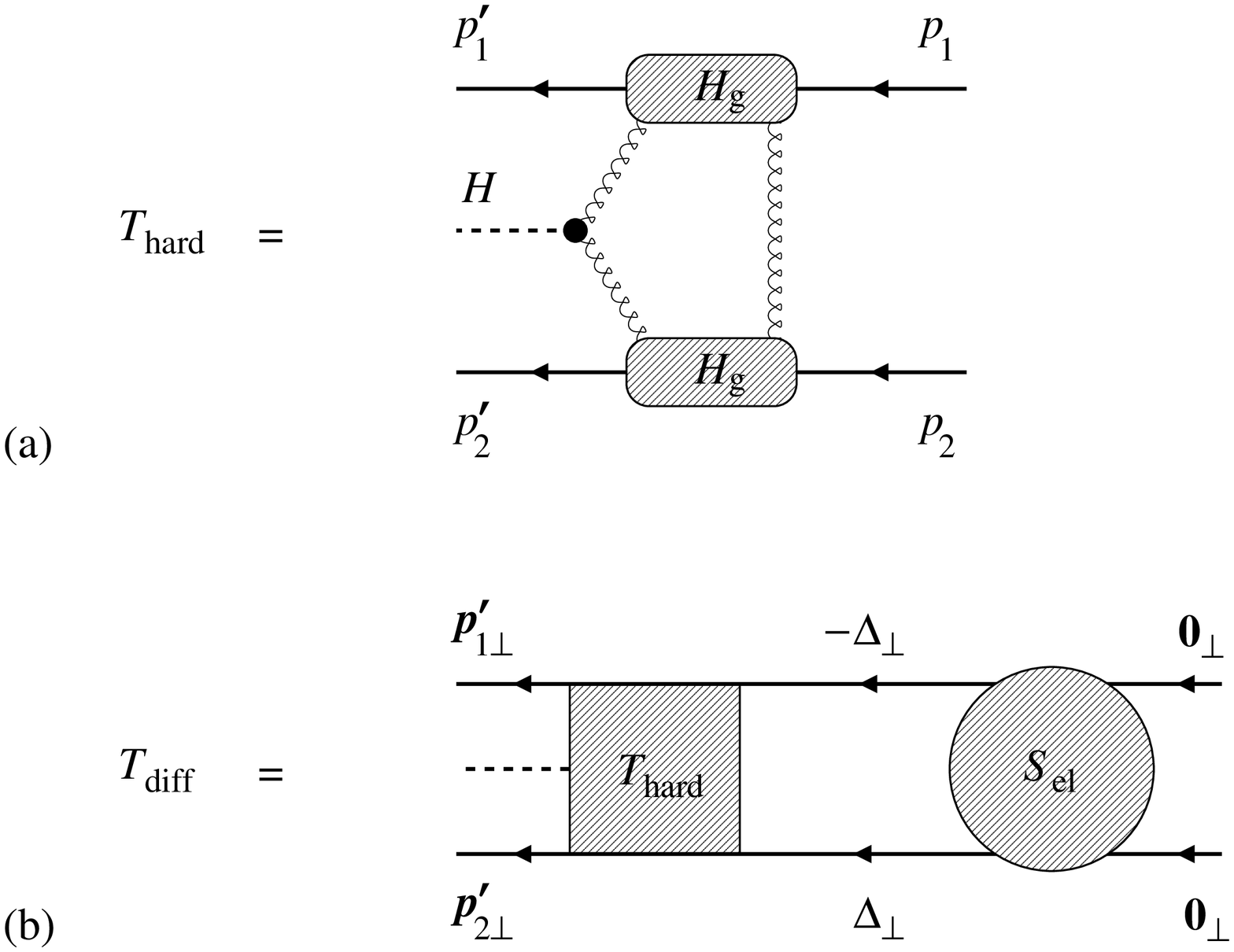}}   
\caption{\label{fig:diag}}
\end{figure}

Following Ref.\cite{RGS}, the amplitude for the hard scattering
process in exclusive diffraction (see Fig.~\ref{fig:diag}a) can be 
represented as
\begin{equation}
T_{\text{hard}}({\bf p}_{1\perp}^\prime, {\bf p}_{2\perp}^\prime; 
{\bf p}_{1\perp}, {\bf p}_{2\perp})  \; = \; \kappa \;
H_g (x_1,\xi_1,t_1) \; H_g (x_2,\xi_2, t_2),
\end{equation}
where $\kappa$ represents the overall normalization,
$H_g$ is the gluon GPD of the proton, $\xi_{1, 2}$ the longitudinal
momentum loss of the protons, $x_{1, 2} \approx \xi_{1, 2}$, 
and $t_1 \approx -({\bf p}_{1\perp}^\prime - {\bf p}_{1\perp})^2$ 
\textit{etc.} the invariant momentum transfers to the protons. 
The gluon GPD refers to a scale, $Q^2$, of the order
$\Lambda_{\rm QCD}^2 \ll Q^2 \ll M_H^2 \ll s$; 
by QCD evolution it can be related to the ``diagonal'' GPD at a 
lower scale. The amplitude for the diffractive process 
(\ref{exclusive_diffraction}) (see Fig.~\ref{fig:diag}b) 
is then given by (we suppress all arguments except the proton 
transverse momenta)
\begin{eqnarray}
T_{\rm diff}({\bf p}_{1\perp}^\prime,{\bf p}_{2\perp}^\prime )
&=& \int \frac{d^2\Delta_\perp}{(2\pi)^2} \;
T_{\rm hard} \left({\bf p}_{1\perp}^\prime,{\bf p}_{2\perp}^\prime;
\Delta_\perp,-\Delta_\perp \right) \;
S_{\rm el}(s, \Delta_\perp ) .
\label{T_diff}
\end{eqnarray}
Here $S_{\rm el}$ is the $S$--matrix of $pp$ elastic scattering,
which can be expressed in terms of the elastic scattering amplitude,
$T_{\rm el}(s, t)$, {\it viz.}\ its profile function in impact parameter
representation, $\Gamma (s, {\bf b})$, as
\begin{eqnarray}
S_{\rm el}(s, \Delta_\perp ) &=& (2\pi)^2 \; \delta^{(2)}(\Delta_\perp)
\; + \; (4\pi i/s)\, T_{\rm el}(s, t = -\Delta_\perp^2 ) 
\label{T_el} 
\\
&=& \int d^2 b \; e^{-i\Delta_\perp\cdot{\bf b}} 
\left[1 - \Gamma(s,{\bf b})\right] .
\label{Gamma}
\end{eqnarray}
Equation~(\ref{T_diff}) expresses the basic idea of RGS ---
the modification of the original hard scattering amplitude by soft
elastic interactions. It actually describes an interference 
phenomenon: The diffractive amplitude
is the sum of the original hard amplitude 
[``$1$'' in Eq.~(\ref{Gamma})] and the amplitude modified by
elastic scattering [``$\Gamma$'' in Eq.~(\ref{Gamma})]. 
Eq.~(\ref{T_diff}) reproduces the ``geometric''
expression for the RGS probability derived heuristically in
Refs.\cite{Frankfurt:2004kn,Frankfurt:2005mc}. It is based on 
the approximation of independence of hard and soft interactions. 
In Ref.\cite{RGS} we show how correlations between hard and soft 
interactions ({\it e.g.}\ due to scattering from the proton's long--range 
pion field, or to parton clustering in ``constituent quarks'') 
can modify this simple result.

The profile function of the $pp$ elastic amplitude has been
determined from fits to the $pp$ elastic and total cross section data.
At TeV energies, it is expected that it approaches the black disk limit 
(BDL) at small impact parameters \cite{Frankfurt:2004kn,Frankfurt:2005mc},
$\Gamma(s, b) \rightarrow 1$ for $b<b_0(s)$.
For illustrative purposes, we make a simple gaussian ansatz
(although it satisfies the BDL only as $b\rightarrow 0$),
$\Gamma(s,{\bf b}) = \exp\left\{-{\bf b}^2/\left[ 2 B(s) \right]\right\}$,
where $B = 21.8\, {\rm GeV}^{-2}$ at $\sqrt{s} = 14\, {\rm TeV}$
from extrapolation of fits to the present data. 
The $t$--dependence of the gluon GPD has been
extracted from measurements of exclusive $J/\psi$ 
photo/electroproduction\cite{Aktas:2005xu}; 
see Ref.\cite{Frankfurt:2005mc} for an overview. 
We parametrize it by a simple exponential, 
$H_g(x, \xi, t) \propto \exp(B_g t/2)$, with
$B_g = 3.24\, {\rm GeV}^{-2}$ in the $x$-- and $Q^2$ 
range relevant to production of a system with $
M_H \approx 100 \, {\rm GeV}$ at the LHC; see Ref.\cite{RGS} for details. 
Notice that the average squared transverse radius of the distribution of
hard gluons is much smaller than the squared radius of soft
interactions: $2 B_g \ll B$.

%
%
\begin{figure}[t]
\epsfxsize=1.0\textwidth
\epsfbox{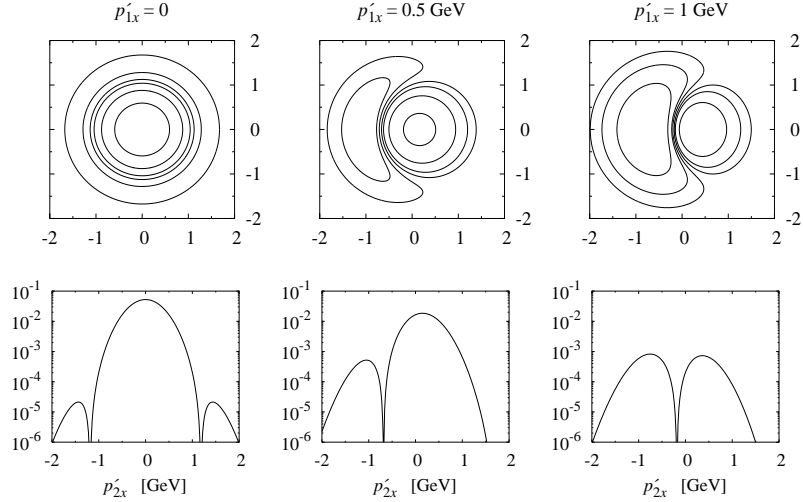}
\caption[]{Transverse momentum dependence of the cross section 
for exclusive double--gap diffraction (\ref{exclusive_diffraction}).
The plots show $|T_{\rm diff}|^2$ (with $\kappa = 1$) as a function of
${\bf p}_{2\perp}^\prime$, for three values of
${\bf p}_{1\perp}^\prime=(p_{1x}^\prime, 0)$.
The top plots are 2--dimensional log--scale contour 
plots in ${\bf p}_{2\perp}^\prime$;
the bottom plots show the distributions for 
${\bf p}_{2\perp}^\prime=(p_{2x}^\prime, 0)$.
\label{fig:Diffraction}}
\end{figure}
With the parametrizations for $\Gamma$ and $H_g$ we can explore the
transverse momentum dependence of the diffractive amplitude
(\ref{T_diff}). Figure~\ref{fig:Diffraction} shows
$|T_{\rm diff}|^2$ as a function of ${\bf p}_{\perp 2}'$, for 
fixed values of ${\bf p}_{1\perp}^\prime$. The pattern can be
understood as the wave generated by $T_{\rm hard}$ diffracting 
off the ``hole'' formed by the surface--dominated soft interaction
profile, $1 - \Gamma (s, {\bf b})$. By measuring this dependence 
in dijet production one can perform detailed 
tests of the diffractive reaction mechanism, and even extract information 
about the gluon GPD in the proton. Such studies appear to be feasible
with the proposed forward detectors at the LHC. In Higgs production, 
the diffraction pattern serves as an indicator of the quantum numbers
of the produced system; in $0^+$ production (shown here) 
the protons preferentially emerge at the same side of the beam, 
in $0^-$ production at a $90^\circ$ angle.

{\footnotesize
Notice: Authored by Jefferson Science Associates, LLC under U.S.\ DOE
Contract No.~DE-AC05-06OR23177. The U.S.\ Government retains a
non-exclusive, paid-up, irrevocable, world-wide license to publish or
reproduce this manuscript for U.S.\ Government purposes.
Supported by other DOE contracts and the 
Binational Science Foundation.
}

\end{document}